\documentclass [12pt]{article}
\usepackage{graphicx,amssymb,amsfonts,latexsym,amsmath,amsthm,times}
\usepackage{epsfig}
\usepackage{fancyhdr}
\usepackage{color}
\usepackage[english]{babel}
\numberwithin{equation}{section}

\def \R {{\mathbb R}}

\def \Tset {{\mathbb T}}
\def \Zset {{\mathbb Z}}
\def \d {{\mathrm d}}
\def \x {{\mathrm {\bf x}}}
\def \V {{\mathrm {\bf V}}}
\def \A {{\mathcal A}}
\def \S {{\mathcal S}}
\newcommand{\ra}{\rightarrow}
\newcommand{\be}{\begin{equation}}
\newcommand{\ee}{\end{equation}}

\newcommand{\ar}{\mbox{$\alpha$}}
\newcommand{\gam}{\mbox{$\gamma$}}

\begin{document}

\footnotesize {\flushleft \mbox{\bf \textit{}}}
 \\
\mbox{\textit{{\bf }}}

\thispagestyle{plain}

\vspace*{2cm} \normalsize \centerline{\Large \bf Turbulent Flame Speeds of G-equation Models}
\centerline{\Large \bf in Unsteady Cellular Flows}

\vspace*{1cm}

\centerline{\bf Yu-Yu Liu $^a$\footnote{Corresponding
author. E-mail: yuyul@mail.ncku.edu.tw}, Jack Xin $^b$ \footnote{Email: jxin@math.uci.edu}
and Yifeng Yu $^b$ \footnote{Email: yyu1@math.uci.edu.}}

\vspace*{0.5cm}

\centerline{$^a$ Department of Mathematics, National Cheng Kung University, Tainan 70101, Taiwan}

\centerline{$^{b}$ Department of Mathematics, University of California, Irvine, CA 92697, USA}


\vspace*{1cm}

\noindent {\bf Abstract.}
We perform a computationl study of front speeds of G-equation models in time dependent 
cellular flows. The G-equations arise in premixed turbulent combustion, and are Hamilton-Jacobi type level set 
partial differential equations (PDEs). The curvature-strain G equations are also non-convex 
with degenerate diffusion. The computation is 
based on monotone finite difference discretization and 
weighted essentially nonoscillatory (WENO) methods. We found that the 
large time front speeds lock into 
the frequency of time periodic cellular flows in curvature-strain G-equations similar to 
what occurs in the basic inviscid G-equation. However, such frequency locking phenomenon disappears in 
viscous G-equation, and in the inviscid G-equation if time periodic oscillation of the 
cellular flow is replaced by time stochastic oscillation.

\vspace*{0.5cm}

\noindent {\bf Key words:} G-equations, front speed computation, cellular flows, frequency locking.

\noindent {\bf AMS subject classification:} 70H20, 76F25, 76M20


\vspace*{1cm}

\setcounter{equation}{0}
\section{Introduction}

Front propagation in turbulent combustion is a nonlinear multiscale dynamical process \cite{P00,X00,X09}. 
In gasoline engine, the burning velocity determines the engine efficiency and the combustion cycle. 
Hence to measure and study the flame propagation speed is a fundamental issue in combustion experiments 
and theory. The comprehensive governing equations for flame propagation involve 
Navier-Stokes equations coupled with transport equations, 
which express  laws of the fluid dynamics and the chemical reactions respectively. 
Simplified geometric models, such as the level set G-equations \cite{W85,P00},  
are often more efficient in improving our understanding of such complex phenomenon.

In level set approach \cite{OF02}, the so-called G-equation in turbulent combustion is:
\be\label{Gi}
G_t+\V(\x,t)\!\cdot\! DG+s_L|DG|=0,
\ee
which describes the motion of the flame front at a prescribed flow velocity $V(\x,t)$ and 
a constant speed $s_L$ along the normal direction of the flame front called the "laminar flame speed". 
When the flame front is planar and the flow velocity is at rest, the front propagates at the speed $s_L$. 
If the flow is in motion, the front is wrinkled in time and eventually propagates at an 
asymptotic speed $s_T$ in each specified direction, which is called the "turbulent flame speed". 
Through the framework of homogenization theory, the existence of the turbulent flame speed of (\ref{Gi}) 
has been rigorously established for periodic flows \cite{XY10,CNS11} and random flows \cite{NN11,CS12}.

As fluid turbulence is known to cause corrugations of flames, additional modeling terms are often  
incorporated into the basic G-equation (\ref{Gi}) on physical grounds \cite{P00}. 
We shall study turbulent flame speeds characterized by 
such extended G-equation models involving curvature, diffusion or strain effects. 
The curvature G-equation is:
\be\label{Gc}
G_t+\V(\x,t)\!\cdot\! DG+s_L|DG|=d|DG|\mathrm{div}\left({DG\over|DG|}\right),
\ee
which comes from adding mean curvature term to the basic motion law. If the curvature terms is linearized, we arrive at the viscous G-equation:
\be\label{Gv}
G_t+\V(\x,t)\!\cdot\! DG+s_L|DG|=d\Delta G,
\ee
which is also a model for understanding the numerical diffusion \cite{OF02}. The strain G-equation is:
\be\label{Gs}
G_t+\V(\x,t)\!\cdot\! DG+\left(s_L\!+\!d{DG\!\cdot\! DV\!\cdot\! DG\over|DG^2|}\right)|DG|=d|DG|\mathrm{div}\left({DG\over|DG|}\right)\!.
\ee
The strain term $n\!\cdot\! DV\!\cdot\! n$ comes from the flame surface stretching rate. 
We will give a brief derivation of all G-equation models later.

Our goal is to study turbulent flame speed $s_T$ and its dependence on the velocity field $V(\x,t)$. 
Cellular flows are two dimensional vortical flows with nontrivial geometric streamline 
structures, hence often adopted for mathematical and computational study of $s_T$.  
Let us first consider the steady cellular flow:
$$
\V(\x,t)=A\!\cdot\!(-\sin(2\pi x)\cos(2\pi y),
\cos(2\pi x)\sin(2\pi y)).
$$
which is a two dimensional periodic incompressible flow with parameter $A$ as the flow intensity. 
The streamline of the cellular flow consist of a periodic array of hyperbolic 
(saddle points, separatrices) and elliptic (vortical) regions. 
For inviscid G-equation, \cite{O01,ACVV02} showed that $s_T$ is enhanced by 
the cellular flow with growth rate $s_T=O(A/\log A)$. 
For viscous G-equation, it is proved in \cite{LXY11} that the diffusion term 
causes strong front speed bending (saturation), and the growth of $s_T$ slows down to uniform bound: $s_T=O(1)$. 
For curvature and strain G-equations, although the analytical justifications are still lacking, 
numerical results \cite{LXY12} showed that the curvature term causes weaker front 
speed bending than the regular diffusion term. Also the strain rate term causes flame quenching, that is, 
$s_T$ starts to decrease for larger $A$ and eventually drops to zero.

Next, we consider the unsteady (time dependent) cellular flow. 
For inviscid G-equation (\ref{Gi}) and time periodic cellular flow, 
it is observed in \cite{CTVV03} that the front propagation 
may be synchronized by the spatial and temporal periodicity, and $s_T$ is approximately 
frequency times a rational number. Such phenomenon is called "frequency locking". 
The time periodic and random cellular flows are also considered in \cite{NX08} within 
a reaction-diffusion-advection (RDA)
equation model where temporally stochastic perturbations of the cellular flow 
also cause strong resonance (although the number of 
resonance peaks reduces to one).

In this paper, we study turbulent flame speeds in unsteady cellular flows 
for extended G-equation models (\ref{Gc}),(\ref{Gv}),(\ref{Gs}). The organization is as follows. 
In section 2, we give an overview of G-equation models and refer the associated numerical 
schemes based on monotone finite difference methods to \cite{LXY12}. In section 3, we show numerical 
findings. For time periodically shifted cellular flows, we found that frequency locking of the front propagation 
persists in curvature and strain G-equations, and the locking phenomenon is robust 
with respect to the coefficient $d$ of the curvature or strain term. 
On the other hand, it disappears in the viscous G-equation (\ref{Gv}), and in time
random cellular flows where $s_T$ decreases in the  
oscillation ``frequency'' (the reciprocal of correlation length). 
For time periodically amplified cellular flows, all results are similar except that 
if the amplitude oscillation frequency is too high then the front becomes insensitive to the cellular flow. 
We end with concluding remarks in section 4, and acknowledgements in section 5. 


\vspace*{0.5cm}
\setcounter{equation}{0}
\section{G-equation Models}

In the thin reaction zone regime and the corrugated flamelet regime of premixed turbulent combustion \cite{P00}, the flame front is modeled by a level set function $\{G(\x,t)=0\}$, which is the interface between the unburned fuel $\{G<0\}$ and burned fuel $\{G>0\}$. The trajectory of a particle $\x(t)$ on the interface is given by a velocity field and a laminar speed:
\be\label{Law}
{\d \x\over\d t}=\V(\x,t)+s_Ln,
\ee
where the positive constant $s_L$ is called the laminar flame speed and $n=DG/|DG|$ is the 
normal velocity ($D$: spatial gradient). In terms of $G$, the motion law  (\ref{Law}) gives the inviscid G-equation (\ref{Gi}). To take account of the effect of flame stretching, the surface stretch rate may be added as a first order correction term on laminar flame speed:
$$\begin{array}{c}
\hat{s}_L=s_L-d(\kappa+\S),\medskip\\
\kappa=\mathrm{div}(n),
\ \mathcal S=-n\!\cdot\! DV\!\cdot\! n.
\end{array}$$
Here $d$ is called the Markstein diffusive number, $\kappa$ is the mean curvature and $\mathcal S$ is called the strain rate. Replacing $s_L$ by $\hat{s}_L$ in the motion law (\ref{Law}), we obtain the strain G-equation (\ref{Gs}).
If we include the motion by curvature only, we have the curvature G-equation (\ref{Gc}). If the curvature term is further linearized, we arrive at the viscous G-equation (\ref{Gv}).

Suppose the flame front propagates in $x$-direction in two dimensional space ($\x=(x,y)$). Consider the stripe domain $\R\times[0,1]$, and the burned region at time $t$ is $\{\x\in\R\times[0,1]:G(\x,t)<0\}$. Denote $\A(t)$ the area that the burned region has invaded during time interval $(0,t)$:
$$
\A(t)=\int_{\R\times[0,1]}\chi_{\{G(\x,t)<0\}} -\chi_{\{G(\x,0)<0\}}\d\x.
$$
($\chi$: indicator function.) We refer $\A\rq{}(t)$ as the instantaneous burning velocity. Then turbulent flame speed is the large time linear growth rate of $\A(t)$ or the average of $\A\rq{}(t)$:
\be\label{St}
s_T=\lim_{t\ra\infty}{\A(t)\over t}=\lim_{t\ra\infty}{1\over t}\int_0^t\A'(t)dt.
\ee
In numerical simulation, consider G-equation with initial data $G(\x,0)=x$. If $\V(\x,t)$ is spatially periodic, we can reduce the computational domain to $[0,1]^2$ by imposing the affine periodic condition:
\be\label{IVP}
\left\{\begin{array}{ll}
G_t+\V(\x,t)\!\cdot\! DG+s_L|DG|=0 & \mbox{in}\ \Tset^2\times (0,\infty)
\\
G(\x,0)=x& \mbox{on}\ \Tset^2\times \{t=0\}
\\
G(\x+k{\bf e}_1,t)=G(\x,t)+k& \x\in [0,1]^2, \ k\in\Zset
\end{array}\right..
\ee
Based on the framework of finite difference computation of Hamilton-Jacobi equations \cite{OF02}, the first order derivatives (flow velocity, laminar speed and strain terms) are discretized as monotone numerical Hamiltonian and approximated by weighted essentially nonoscillatory (WENO) scheme. The second order derivatives (curvature and diffusion terms) are approximated by central differencing. For the explicit time step discretization, total variation diminishing Runge-Kutta (TVD RK) scheme is used, and the time step size restriction is given by the CFL condition. See \cite{LXY12} for details of the numerical schemes. We compute (\ref{IVP}) with spatial grid size $200\times200$.

\vspace*{0.5cm}
\setcounter{equation}{0}
\section{Numerical Results}
Our first set of numerical results is on the time periodically shifted cellular flow:
$$
\V(\x,t)=A\!\cdot\!(-\sin(2\pi x+B\sin(2\pi\omega t))\cos(2\pi y),
\cos(2\pi x+B\sin(2\pi\omega t))\sin(2\pi y)).
$$
The flow oscillates temporally in the $x$-direction with amplitude $B$ and frequency $\omega$. 
We fix $s_L=1$, $A=4$, $B=1$ and choose different values for $d$ to see the effect of the curvature, 
diffusion and strain terms. Note that if $d=0$ then all G-equations are identical to inviscid G-equation, 
and we choose $d$ small enough to avoid the strain term overpowering $s_L$.

\begin{figure}\center
\includegraphics[width=0.48\textwidth]{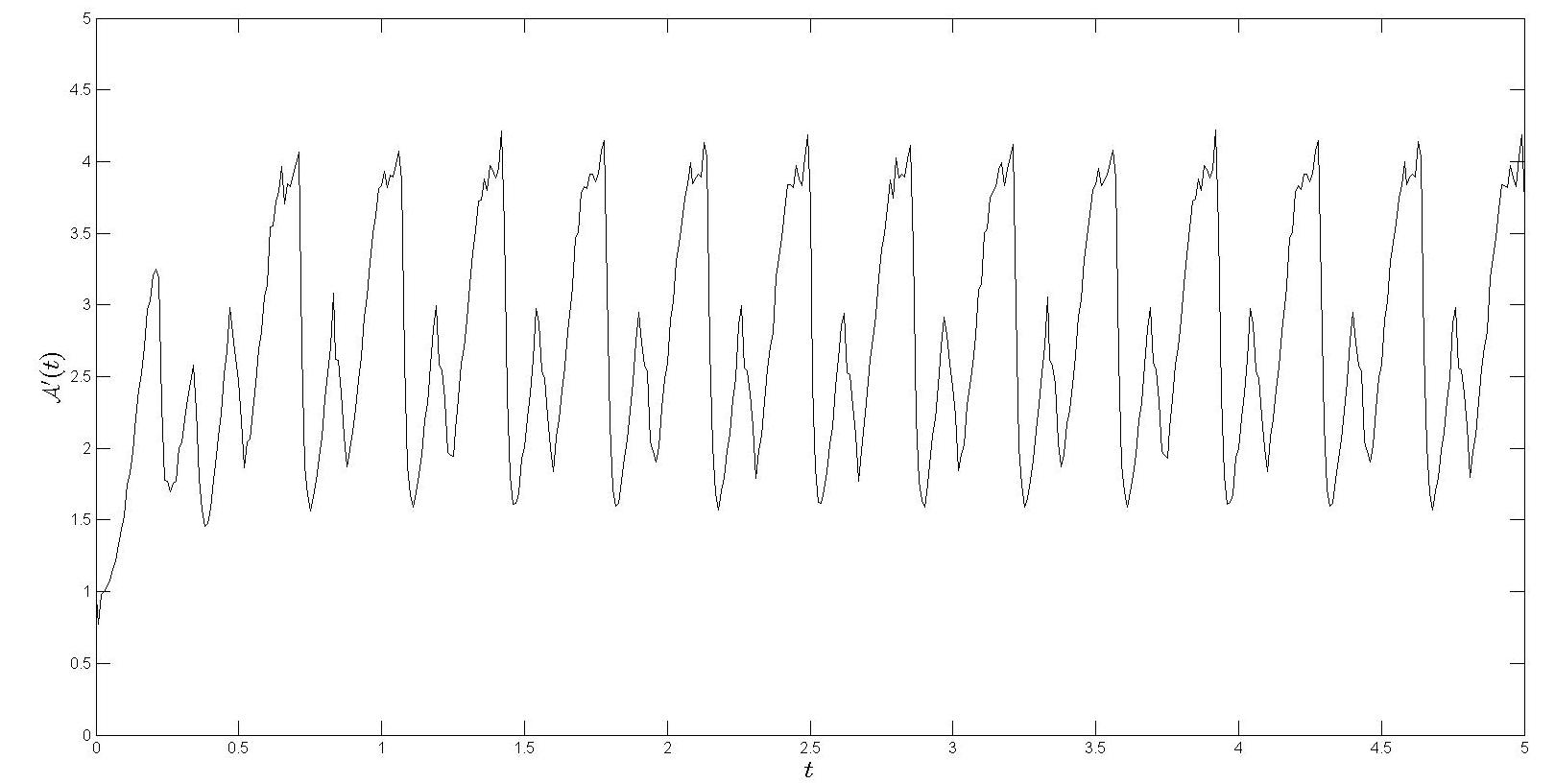}
\includegraphics[width=0.48\textwidth]{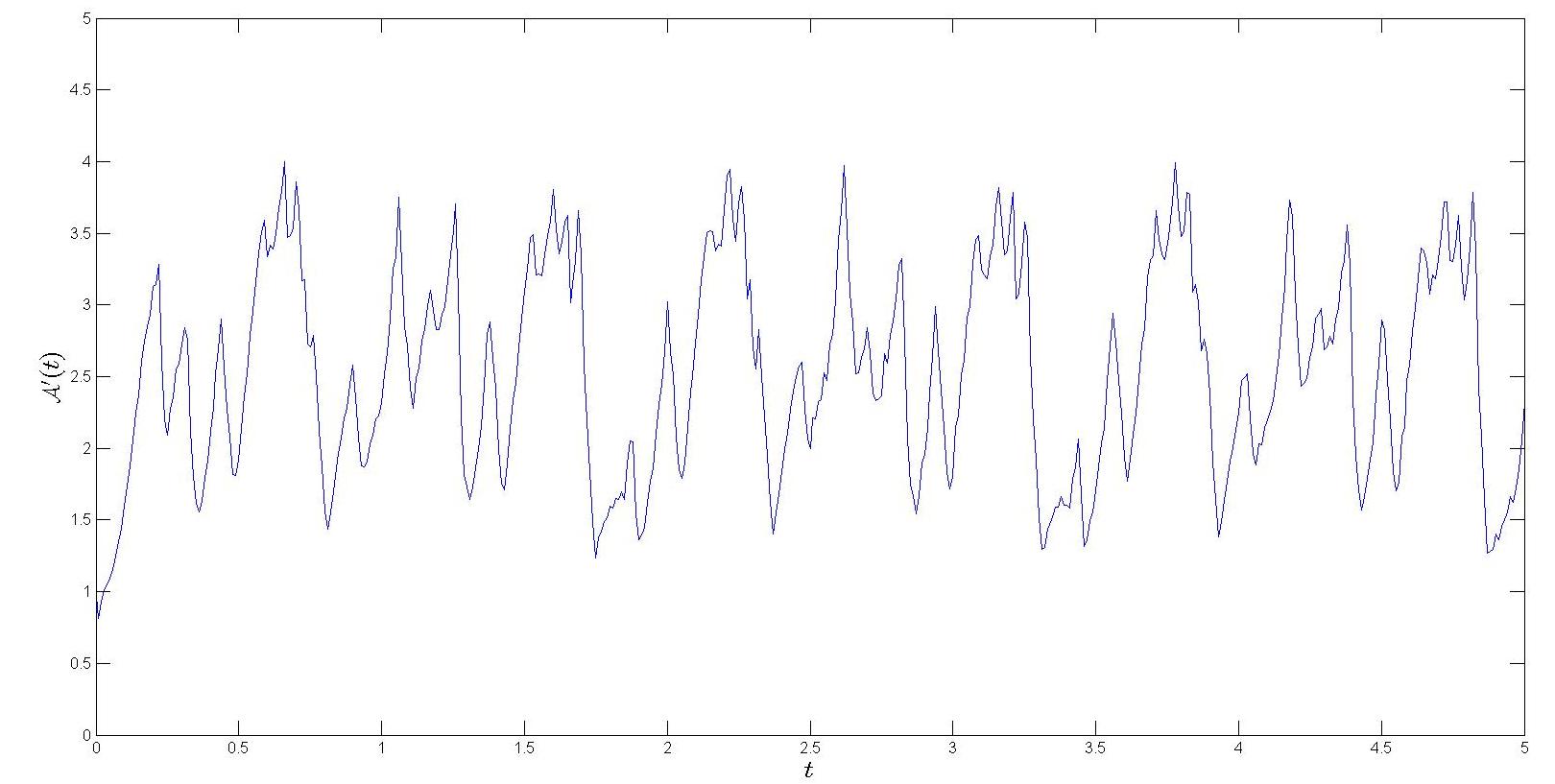}
\caption{Plots of $\A\rq{}(t)$ for inviscid G-equation with time periodically shifted 
cellular flow, $s_L=1$, $A=4$, $B=1$. Right: $\omega=2.8$. Left: $\omega=3.2$.}
\label{A(t)}
\end{figure}

Figure \ref{A(t)} shows the plot of $\A\rq{}(t)$ for inviscid G-equation. 
For $\omega=2.8$, $\A\rq{}(t)$ evolves into a periodic function after a transient time. 
In this case frequency locking occurs, and we can simply compute $s_T$ by taking the average of $\A\rq{}(t)$ over a few periodic intervals. For  $\omega=3.2$, $\A'(t)$ remains chaotic over time, and $s_T$ is evaluated by taking average of $\A'(t)$ over a large time interval as (\ref{St}).

The frequency locking effect can be understood as follows. In the transient time, the front propagation structure is synchronized with the temporal and spatial periodicity of the cellular flow and evolves into a periodic steady state. During a multiple of the oscillatory period $\Delta t=M\cdot 1/\omega$, the flame front translates through a multiple of the cell $\Delta\A(t)=N$. ($N,M$: integers.)
Then the turbulent flame speed is:
\be\label{Lock}
s_T={\Delta\A(t)\over \Delta t}={N\over M}\!\cdot\!\omega.
\ee

\begin{figure}\center
\includegraphics[width=0.75\textwidth]{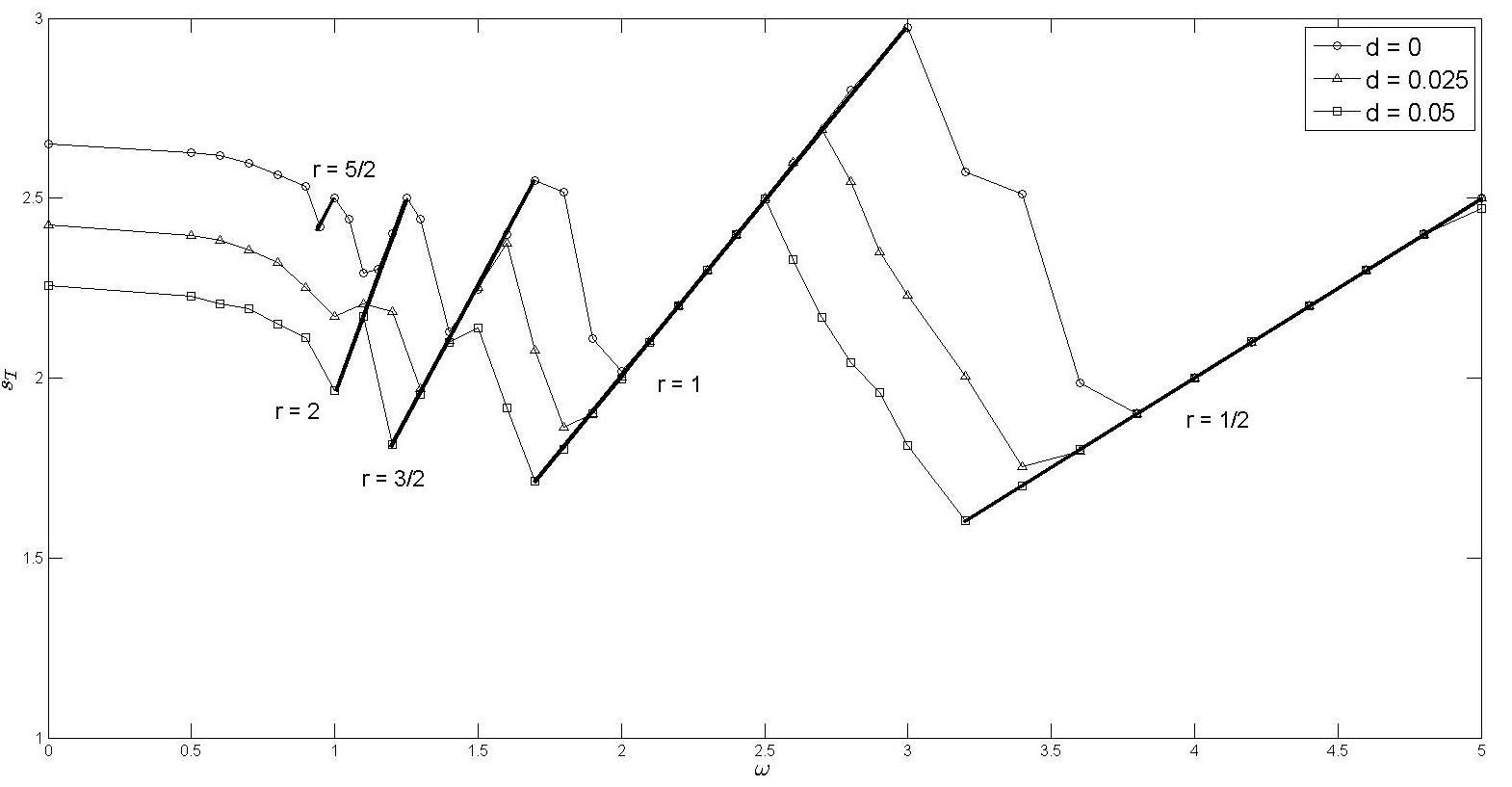}
\includegraphics[width=0.75\textwidth]{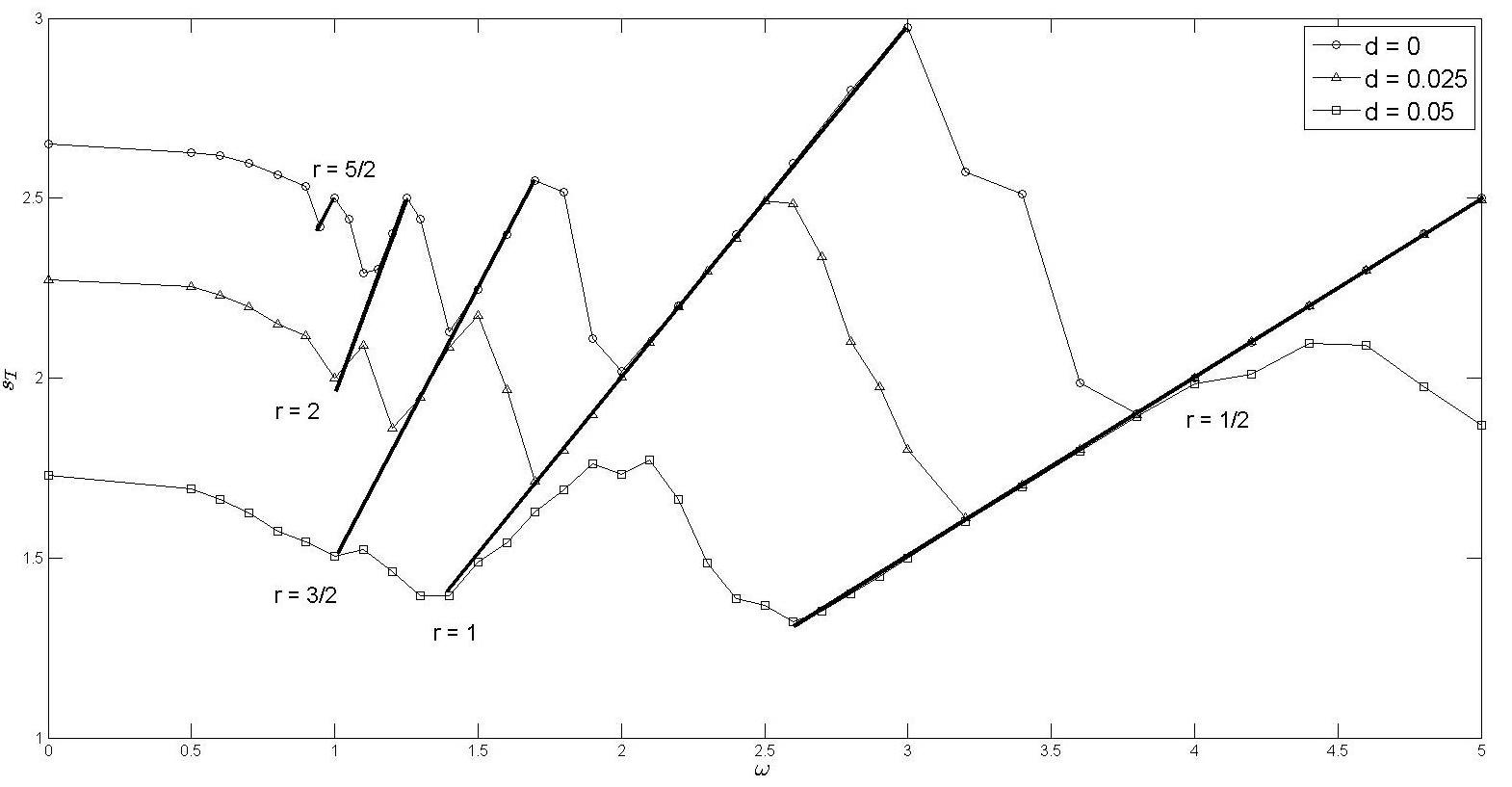}
\includegraphics[width=0.75\textwidth]{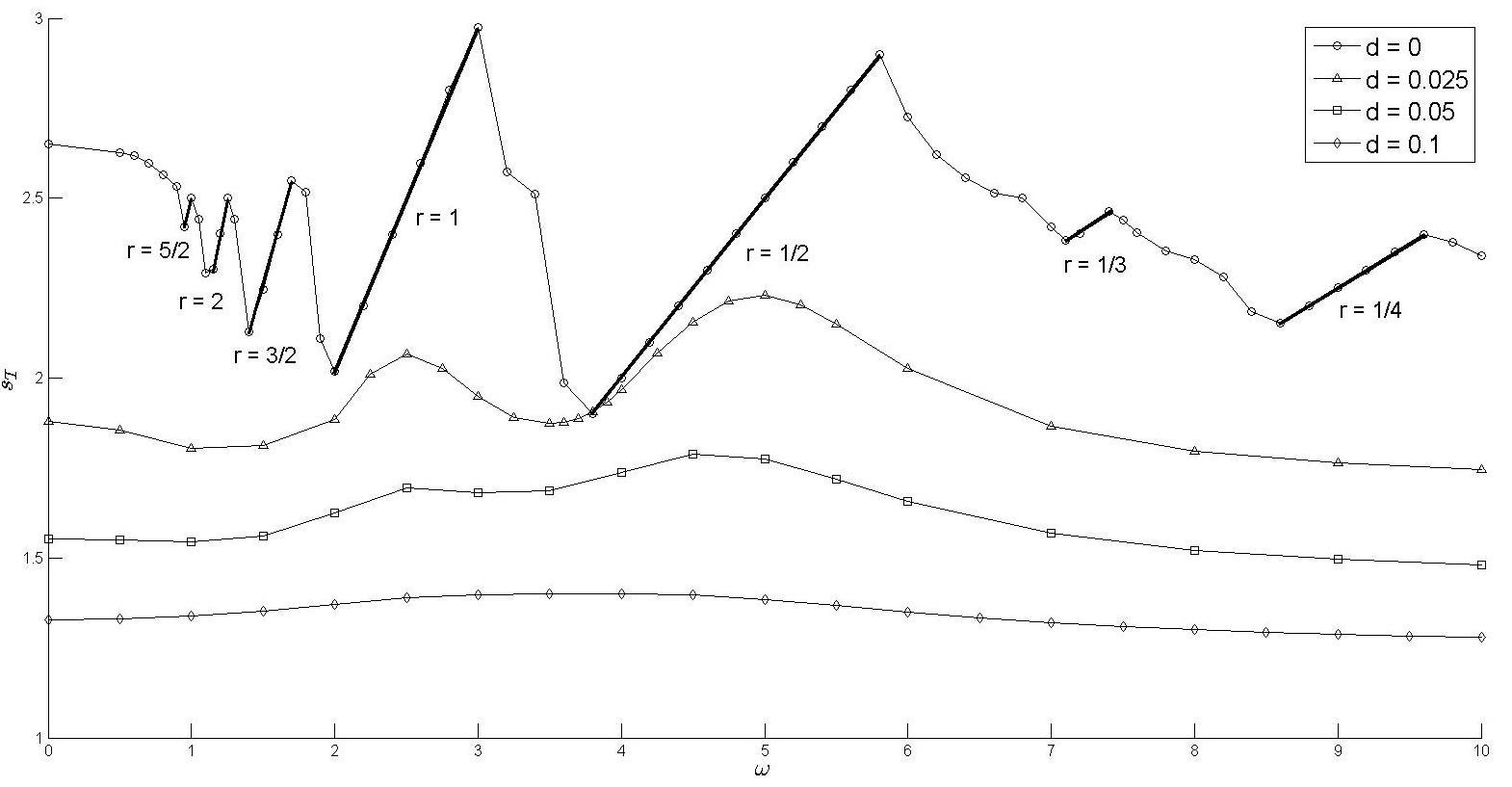}
\caption{Turbulent flame speeds of G-equation models with time periodically shifted cellular flow. 
Above: inviscid/curvature G-equations. Middle: inviscid/strain G-equations. Below: inviscid/viscous G-equations. 
The bold lines indicate the frequency locking pattern $r=s_T/\omega$.}
\label{Shift}
\end{figure}

Figure \ref{Shift} shows the plot of turbulent flame speed as a function of 
the oscillatory frequency $s_T=s_T(\omega)$ for all G-equation models with various $d$. For inviscid, 
curvature and strain G-equations (\ref{Gi}),(\ref{Gc}),(\ref{Gs}), frequency locking is observed. 
Here $s_T(\omega)$ is piecewise linear with 
rational slopes $r=s_T/\omega$. In each linear segment 
the front propagation locks into the same periodicity pattern, 
that is, the same $N,M$ in (\ref{Lock}) and hence same $r$. 
Between the linear segments, the locking effect loses stability and $s_T(\omega)$ decreases. 
Notice that frequency locking phenomenon is also robust with respect to $d$, 
that is, $s_T$ may be the same even as $d$ varies (different curves may coalesce on bold line segments of 
top and middle panels of Figure \ref{Shift}). 
This is different from that in steady cellular flow where
 $s_T$ is strictly decreasing as $d$ increases \cite{LXY12}.

For viscous G-equation, however, no frequency locking appears. 
Instead the dissipation effect causes the level set function to converge to the stationary solution: 
$$G(\x,t)=-s_Tt+x+u(\x,t)$$ 
with $u(\x,t)$ a periodic function in $[0,1]^2\times[0,1/\omega]$. In the plot of $s_T(\omega)$, we 
see two resonant peaks for smaller $d$, then the peaks start to merge and disappear for larger $d$. 

\begin{figure}\center
\includegraphics[width=0.75\textwidth]{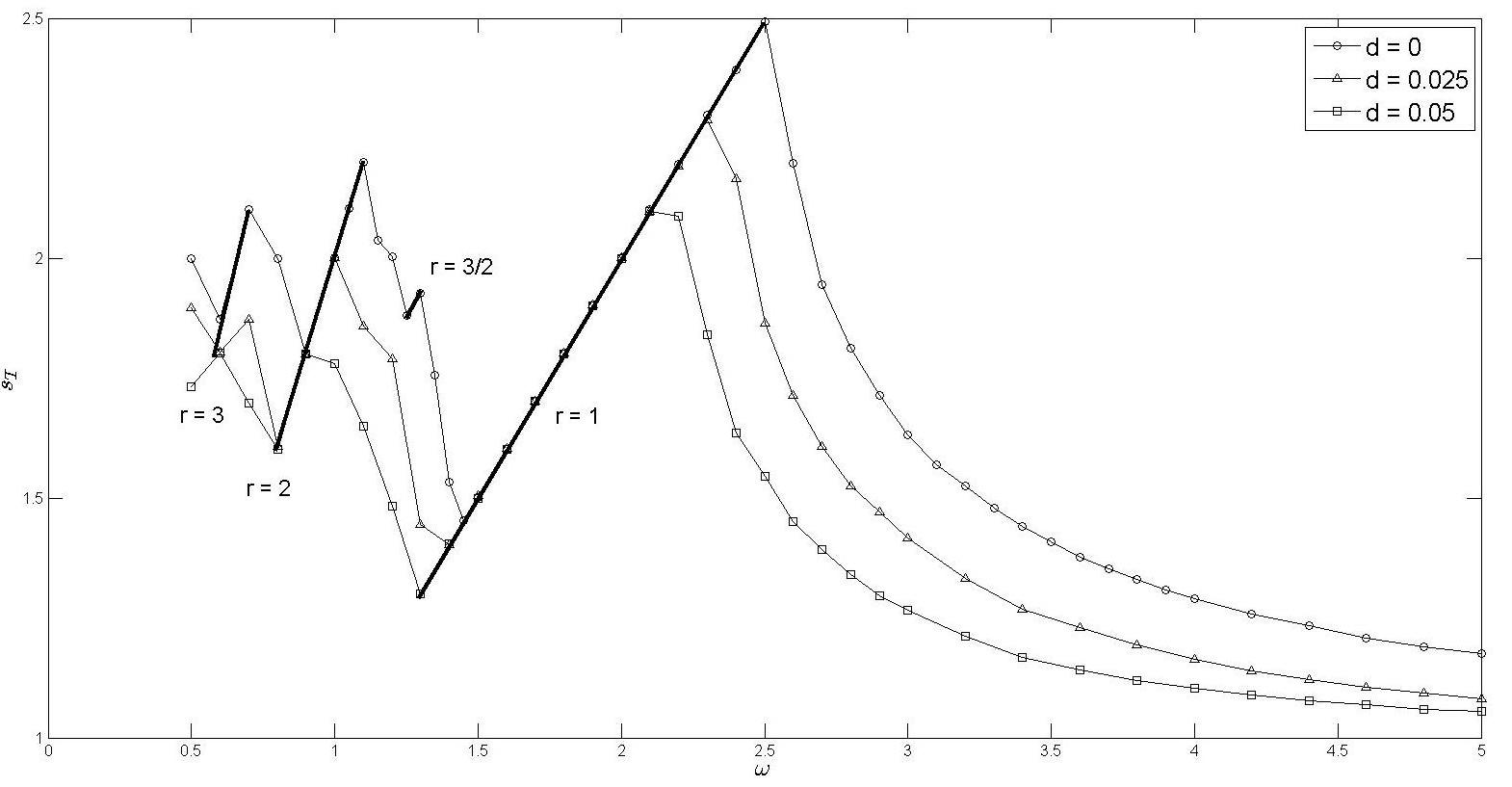}
\includegraphics[width=0.75\textwidth]{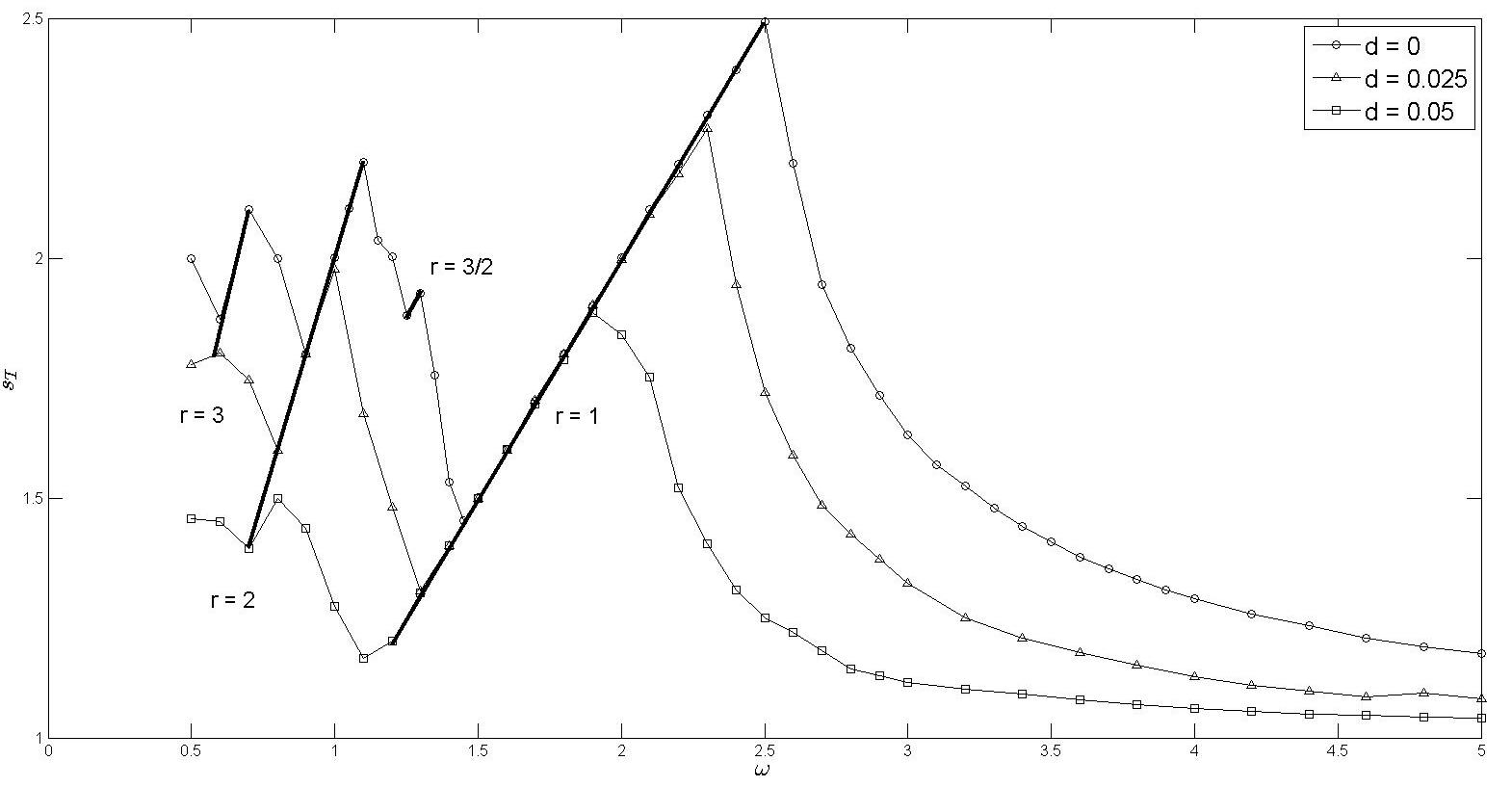}
\includegraphics[width=0.75\textwidth]{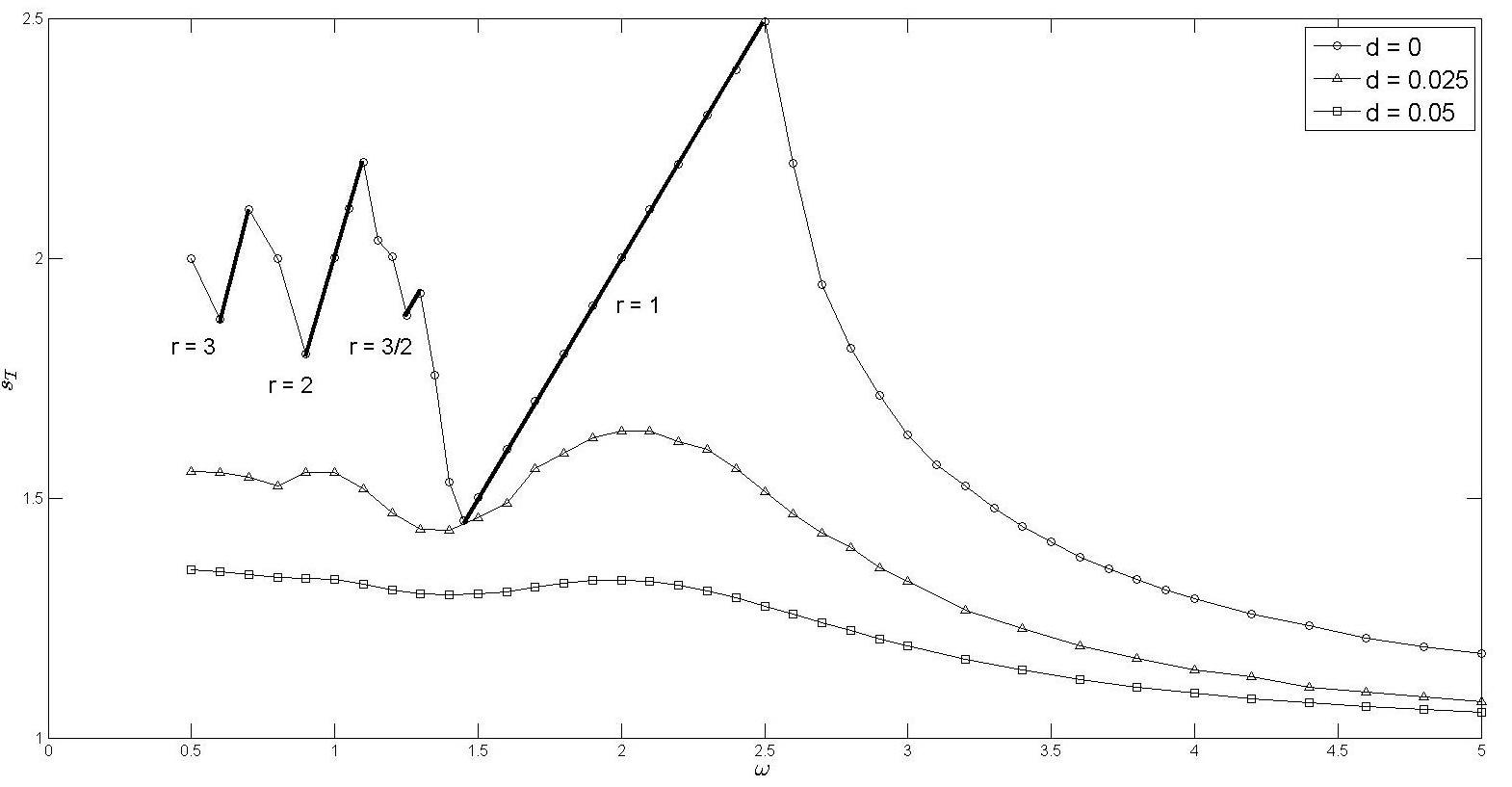}
\caption{Turbulent flame speeds of G-equation models with time periodically amplified cellular flow. 
Above: inviscid/curvature G-equations. Middle: inviscid/strain G-equations. Below: inviscid/viscous G-equations. The bold lines indicate the frequency locking pattern $r=s_T/\omega$.}
\label{Amp}
\end{figure}

Our second set of numerical results is when $\V$ is the time periodically amplified cellular flow:
$$\V(\x,t)=A\sin(2\pi\omega t)\cdot(-\sin(2\pi x)\cos(2\pi y),\cos(2\pi x)\sin(2\pi y)).
$$
Figure \ref{Amp} is the plot $s_T=s_T(\omega)$ for all G-equation models and 
various $d$. When $\omega$ is small, all results are similar to the 
time periodically shifted cellular flow. When $\omega$ is larger, however, 
the direction of the flow changes so fast in time that the flame front cannot be 
really wrinkled, and so remains almost planar. Hence $s_T(\omega)$ is not much   
responsive to the cellular flow patterns, and decreases to $s_L$ as $\omega\ra+\infty$ in all the G-equations.

\begin{figure}\center
\includegraphics[width=0.6\textwidth]{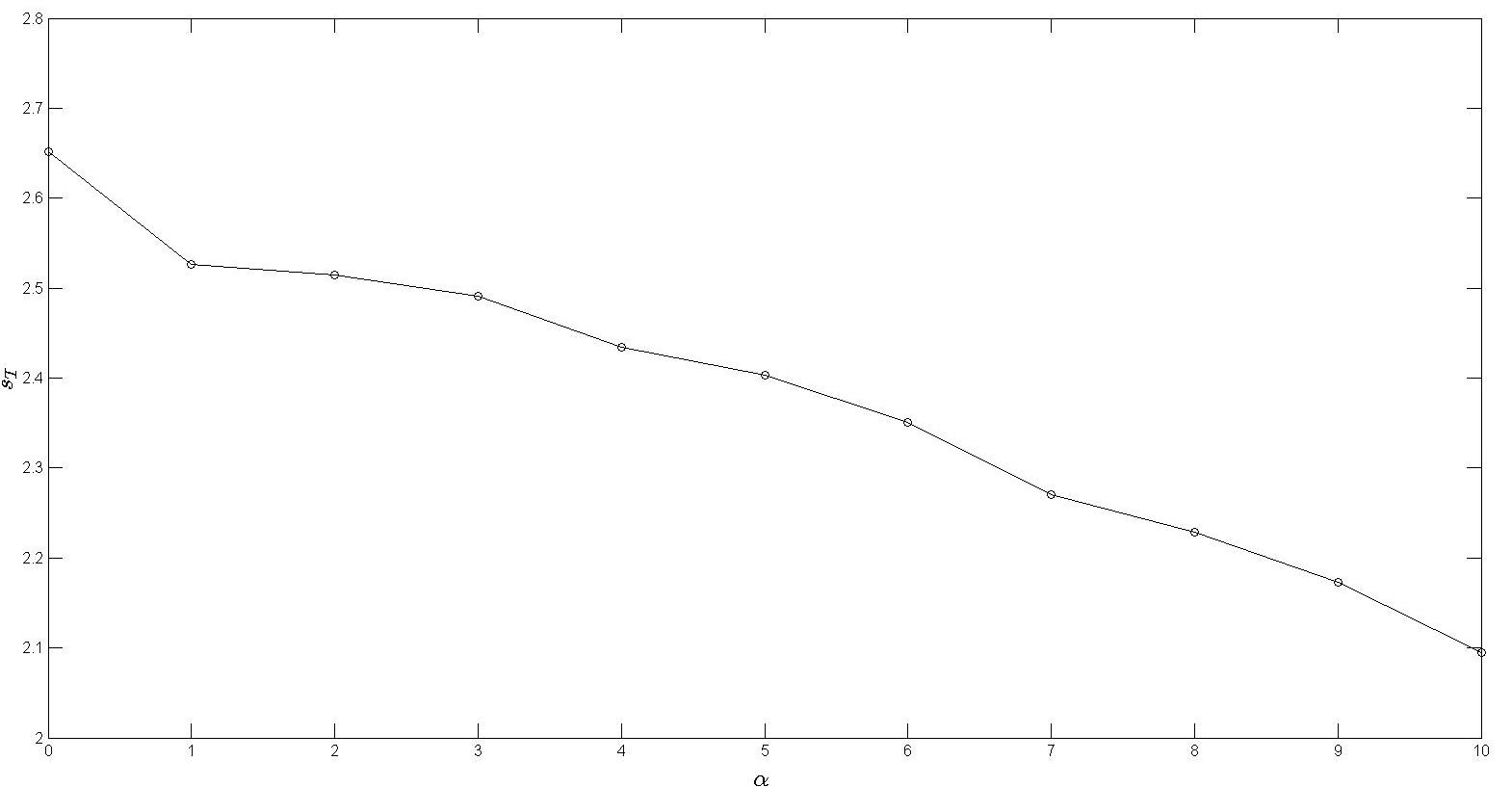}
\caption{Turbulent flame speeds of inviscid G-equation with Ornstein-Uhlenbeck process being  
time oscillatory shift in cellular flow, as a function of $\ar $, the 
reciprocal of the correlation length, or ``oscillation frequency''.}
\label{OU}
\end{figure}

Finally we consider the time stochastically shifted cellular flow:
$$
\V(\x,t)=A\!\cdot\!(-\sin(2\pi x+\gam(t))\cos(2\pi y),
\cos(2\pi x+\gam(t))\sin(2\pi y)).
$$
Here $\gam(t)$ is the Ornstein-Uhlenbeck process defined by the Ito equation:
$$
\d\gam(t)=-\ar\gam(t)\d t+\beta\d W(t),
$$
where $W(t)$ is the Wiener process, $\ar$ and $\beta$ are positive constants. 
We choose $\beta=\sqrt{2}\ar^{3/4}$ so that the power spectrum is invariant in $\ar$,  
the reciprocal of the correlation length of $\gam (t)$. 
Figure \ref{OU} shows the plot $s_T(\ar)$ for inviscid G-equation. 
We see that $s_T$ always decreases as $\ar$ increases, stochasticity in time 
kills the resonance phenomenon. This is different to that in RDA equation model where one wider resonance peak 
still remains under time stochastic perturbation of cellular flows \cite{NX08}.

\vspace*{0.5cm}
\setcounter{equation}{0}
\section{Concluding Remarks}
We have studied numerically the basic and the curvature-strain dependent G-equation models 
and their corresponding turbulent 
flame speeds in various unsteady cellular flows. Numerical results indicate that the  
frequency locking phenomenon occurs in inviscid, curvature and strain G-equations 
but not in viscous G-equations. It also disappears in time stochastically shifted cellular flows. 
In the future, we plan to study more general unsteady (time periodic or random) flows, including three 
dimensional flows for G-equation models.

\vspace*{0.5cm}

\section*{Acknowledgements}
The work was partially supported by National Science Counsel grants (YL) of Taiwan,
NSF grants DMS-0911277, DMS-1211179 (JX); DMS-0901460, CAREER award DMS-1151919 (YY) of USA.



\begin{thebibliography}{99}

\bibitem{ACVV02}
M. Abel, M. Cencini, D. Vergni, A. Vulpiani.
\textit{Front Speed Enhancement in Cellular Flows.}
Chaos 12 (2002), 481-488.

\bibitem{CTVV03}
M. Cencini, A. Torcini, D. Vergni, A. Vulpiani.
\textit{Thin front propagation in steady and unsteady cellular flows.}
Phys. Fluids 15 (2003), 679-688.

\bibitem{CNS11}
P. Cardaliaguet, J. Nolen, P. E. Souganidis.
\textit{Homogenization and Enhancement for the G-Equation.}
Arch. Rational Mech. Analysis 199 (2011), 527-561.

\bibitem{CS12}
P. Cardaliaguet, P. E. Souganidis.
\textit{Homogenization and Enhancement of the G-equation in Random Environments.}
arXiv:1110.1760, Comm. Pure Appl. Math, to appear.

\bibitem{LXY11}
Y.-Y. Liu, J. Xin, Y. Yu.
\textit{Asymptotics for turbulent flame speeds of the viscous G-equation enhanced by cellular and shear flows.}
Arch. Rational Mech. Analysis 202 (2011), 461-492.

\bibitem{LXY12}
Y.-Y. Liu, J. Xin, Y. Yu.
\textit{A Numerical Study of Turbulent Flame Speeds of Curvature and Strain G-equations in Cellular Flows.}
arXiv:1202.6152, Physica D, to appear.

\bibitem{NN11}
J. Nolen, A. Novikov. 
\textit{Homogenization of the G-equation with incompressible random drift in two dimensions.}
Comm. Math Sci. 9 (2011), 561-582.

\bibitem{NX08}
J. Nolen, J. Xin.
\textit{Computing reactive front speeds in random flows by variational principle.}
Physica D 237 (2008), 3172-3177.

\bibitem{O01}
A. Oberman. Ph.D. thesis, University of Chicago, Chicago, IL, 2001.

\bibitem{OF02}
S. Osher, R. Fedkiw.
Level Set Methods and Dynamic Implicit Surfaces.
Springer-Verlag, New York, NY 2002.

\bibitem{P00}
N. Peters.
Turbulent Combustion.
Cambridge University Press, 2000.

\bibitem{R95}
P.D. Ronny.
\textit{Some Open Issues in Premixed Turbulent Combustion.}
Lecture Notes in Physics 449 (1995), 3-22.

\bibitem{W85}
F. Williams,
\textit{Turbulent Combustion.} 
The Mathematics of Combustion (J. Buckmaster, ed.), SIAM, Philadelphia (1985) 97-131.

\bibitem{X00}
J. Xin,
\textit{Front Propagation in Heterogeneous Media.}
SIAM Review 42 (2000), 161-230.

\bibitem{X09}
J. Xin.
An Introduction to Fronts in Random Media. 
Surveys and Tutorials in the Applied Mathematical Sciences 5, Springer, 2009.

\bibitem{XY10}
J. Xin, Y. Yu.
\textit{Periodic Homogenization of Inviscid G-equation for Incompressible Flows.}
Comm. Math Sci. 8 (2010), 1067-1078.
\end{thebibliography}
\end{document}